\documentclass[%
 reprint,
superscriptaddress,
 amsmath,amssymb,
 aps,
 floatfix,
]{revtex4-2}

\usepackage{graphicx}% Include figure files
\usepackage{dcolumn}% Align table columns on decimal point
\usepackage{bm}% bold math
\usepackage[mathlines]{lineno}% Enable numbering of text and display math
\begin{document}

% \preprint{APS/123-QED}

\title{Dynamic cost allocation allows network-forming forager to switch between search strategies}% Force line breaks with \\

\author{Lisa Schick}
\affiliation{%
 Technical University of Munich, TUM School of Natural Sciences, Department of Bioscience, Center for Protein Assemblies (CPA), Germany
}%
\author{Mirna Kramar}
\affiliation{Institute Curie, UMR168, Paris, France}%Lines break automatically or can be forced with \\
\author{Karen Alim}%
 \email{k.alim@tum.de}
\affiliation{%
 Technical University of Munich,  TUM School of Natural Sciences, Department of Bioscience, Center for Protein Assemblies (CPA), Germany
}%

\date{\today}% It is always \today, today,

\begin{abstract}
Network-forming organisms, like fungi and slime molds, dynamically reorganize their networks during foraging. The resulting re-routing of resource flows within the organism's network can significantly impact local ecosystems. In current analysis limitations stem from a focus on single-time-point morphology, hindering understanding of continuous dynamics and underlying constraints. Here, we study ongoing network reorganization in the foraging slime mold \textit{Physarum~polycephalum}, identifying three distinct states with varying morphology and migration velocity. We estimate the energetic cost of each state and find a trade-off between building and transport costs within the morphological variability, facilitating different search strategies. Adaptation of state population to the environment suggests that diverse network morphologies support varied foraging strategies, though constrained by associated costs. Our findings provide insights for evaluating the impact of resource flow re-routing in changing ecosystems.
\end{abstract}


\maketitle
\thispagestyle{empty}
%\tableofcontents
\section{\label{sec:Introduction}Introduction}
%\subsection{Mycelia fungi as key transport network - how are network morphology and function coupled}
%--------FIGURE--------
\begin{figure*}[t]
\centering
\includegraphics[width=17.8cm]{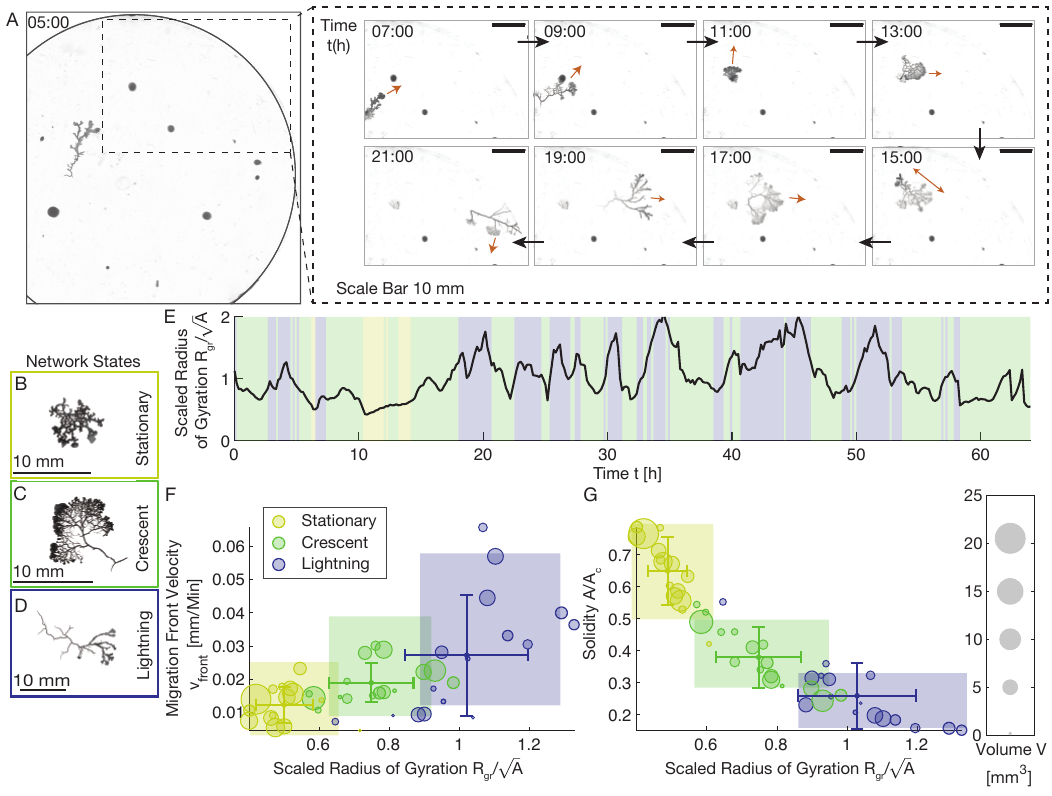}% Here is how to import EPS art %width=17.8cm
\caption{\label{fig:phaseplot} \textit{P.~polycephalum}'s continuous network reorganization is captured by three distinct network states. (A) Network morphology continuously reorganizes during the migration of \textit{P.~polycephalum} networks over several hours (brown arrows indicating migration direction), presented here for the low-time resolution data. Gray spots are heat-killed E.~coli food sources. Network morphology varies among three states (stationary~(B), crescent~(C), lightning~(D)). (E) Time series of the scaled radius of gyration $R_{\mathrm{gr}}/\sqrt{A}$ throughout the experiment for the plasmodium presented in (A) with the states highlighted by the color change in the background. (F,G) Quantification of migration front velocity $v_{\mathrm{front}}$, scaled radius of gyration $R_{\mathrm{gr}}/\sqrt{A}$, and solidity $A/A_c$ show the classification of network states by network dynamics and network morphology. A trained neural network automatically detects network states. Error bars indicate the standard deviation of the distribution of data sets in each state.}
\end{figure*}
%-----------END FIGURE----------

Network-forming foragers, such as fungi and plasmodial slime molds, impress by sheer network size and continuous network reorganization~\citep{fricker_Adaptive_2009, read_Ties_1997, boddy_Saprotrophic_1999,oettmeier_Form_2018,oettmeier_Physarum_2017}. As foragers transport resources throughout their network,  reorganization of their network affects
ecosystems~\citep{hawkins_Mycorrhizal_2023, vantpadje_Quantifying_2021, treseder_Fungal_2015,simard_Mycorrhizal_2012, simard_Carbon_2002}. Both plasmodial slime molds and foraging fungi are strikingly similar in their network reorganization dynamics~\citep{fricker_Adaptive_2009, heaton_Analysis_2012, roper_Mycofluidics_2019} despite their different biological makeup: When foraging, body mass is recycled~\citep{falconer_Biomass_2007, fricker_Network_2007, schenz_Mathematical_2017} as newly formed network fronts enter new territory~\citep{westendorf_Quantitative_2018, baumgarten_Dynamics_2014,glass_Hyphal_2004, bebber_Biological_2007, boddy_Saprotrophic_1999}, thereby changing network morphology~\citep{boddy_Saprotrophic_1999,bebber_Biological_2007, boddy_Saprotrophic_2009, ito_Characterization_2011}. However, how networks reorganize and what costs and functional requirements constrain their reorganization are still being determined.

%\subsection{One task = one morphology}
The central functional role of a foragers' network is to transport resources over long distances governed by network morphology~\citep{nakagaki_Interaction_2000, simonin_Physiological_2012}. 
Evolutionary pressures are expected to constrain network morphology by minimizing transport costs at fixed network building cost~\citep{murray_Physiological_1926}. Depending on network building cost versus transport cost, the optimal network morphology is predicted to be either tree-like (low building cost, but high transport cost) or finely reticulated (high building cost, but low transport cost)~\citep{rocks_Limits_2019,ronellenfitsch_Phenotypes_2019, corson_Fluctuations_2010, katifori_Damage_2010}, even under non-equilibrium conditions~\citep{akita_Experimental_2017}. Within this framework, fungal networks have high building costs, while animal microvasculature has low building costs~\citep{papadopoulos_Comparing_2018}, suggesting species-specific evolutionary constraints resulting in either reticulated or tree-like transport networks. To date, physiological network analysis has focused on network morphology at single points in time, neglecting the impact of continuous network reorganization on network morphology~\citep{bebber_Biological_2007, boddy_Saprotrophic_2009}.

%\subsection{Physarum - model organism}
The plasmodial slime mold \textit{Physarum~polycephalum}~\citep{sauer_Developemental_1982} stands out as a network-forming forager amenable to quantitative network extraction~\citep{bauerle_Spatial_2017,baumgarten_Plasmodial_2010,baumgarten_Functional_2013, fricker_Automated_2017, ito_Characterization_2011, fessel_Physarum_2012} despite its highly dynamic foraging behavior in different environments~\citep{halvorsrud_Growth_1998,vogel_Phenotypic_2015,vogel_Transition_2017, kuroda_Allometry_2015, lee_Novel_2018}. 
\textit{P.~polycephalum} spreads in two-dimensional space (Movie~S1 and~S2, Fig.~\ref{fig:phaseplot}A) at a migration velocity of approximately 0.05~mm/min~\citep{baumgarten_Dynamics_2014,westendorf_Quantitative_2018,kuroda_Allometry_2015}, and its network reorganizes within hours~\citep{marbach_Pruning_2016}.
Rhythmic contractions of the tube walls generate peristaltic fluid flow with a period of 120~s~\citep{wohlfarth-bottermann_Oscillatory_1979}, transporting nutrients, chemical signals, and body mass through the network and into the migration fronts~\citep{nakagaki_Obtaining_2004,alim_Random_2013, alim_Fluid_2018, oettmeier_Lumped_2019, dionne_Active_2024}. However, how body mass reallocation changes the functionality associated with network morphology is unknown.\\
%\subsection{Here we:}
In this study, we investigate how continuous network reorganization of \textit{P.~polycephalum} evolves both over time and as function of nutritious or plain environments. We observe \textit{P.~polycephalum} transitioning continuously between three different network states, regardless of the environment. All states are repeatedly populated over time and differ in network morphology and migration velocity. We calculate the energy costs associated with each network morphology, revealing that energy allocation is a trade-off between transport and building costs among the three states. Upon investigating the function of the different network states, we find that they differ in their search strategies, particularly in how broadly the environment is scanned. Finally, we conclude that the trade-off in energetic costs drives variability in network states and associated search strategies as the population of network states changes in response to the environment. Our observations in \textit{P.~polycephalum} establish the diversity and functionality of the network morphology in network-forming foragers likely impacting resource transport within inaccessible underground foraging fungi.

\section{Results}\label{sec:results}
%%%%%%%%%%%%%%%%%%%%%%%%%%%%%%%%%%%%%%%%%%%%%%%%%%%%%%%%%%%%%%%%%%%%%
% MORPHODYNAMICS
\subsection{\label{sec:part1} Network morphology is dynamic during migration}
%%%%%%%%%%%%%%%%%%%%%%%%%%%%%%%%%%%%%%%%%%%%%%%%%%%%%%%%%%%%%%%%%%%%%
To study the dynamics of network reorganization of \textit{P.~polycephalum} during foraging, we collect time series of bright-field images of 16 individual plasmodial networks at 5-10~minute intervals over 46-60~h (Fig.~\ref{fig:phaseplot}A, Movie~S1 and~S2). The experiments start with well-fed plasmodia, which initially grow without migration. Eventually, after 1-5~h, the plasmodia start migrating on plain 1.5\%~agar to search for new food sources. We present half of the plasmodia with patchy localized food sources of heat-killed E.~coli in varying patterns, enabling 2-3 food encounters for successful foragers. In contrast, the other half migrates on plain agar to evaluate how the environment changes network reorganization.
We prepare networks ranging in size from 2.5-90~mm$^2$ (SI Appendix, Fig.~S1) as the dimensions of our microscope setup limit the observation of more extensive networks, and smaller plasmodia do not form a hierarchical network structure. In all experiments, \textit{P.~polycephalum} undergoes continuous network reorganization on the time scale of minutes to hours (Fig.~\ref{fig:phaseplot}A) varying between static and migratory networks. 
In addition to these low-time resolution data sets, we trim a plasmodial network to prevent it from migrating within a 3~hour time frame of imaging (Movie~S3). We image the continuous network reorganization with 6-second intervals such that we can follow the tube dynamics during the transition between network morphologies with a high-time resolution.
\newline
The observed variability in network morphology is in stark contrast to theoretical work on optimal flow networks~\citep{murray_Physiological_1926, katifori_Damage_2010,corson_Fluctuations_2010, ronellenfitsch_Phenotypes_2019,hu_Adaptation_2013} suggesting a single optimal network state. To capture discovered differences in network shape and migration velocity, we first focus on the low-time resolution data and visually separate experimentally recorded networks into three distinct morphological states (Fig.~\ref{fig:phaseplot}B-D): a stationary state~(B), a crescent state~(C), and a lightning state~(D).
In the stationary state, the plasmodium grows isotropically in size without displacement, while the organism is migratory in both the crescent and lightning states. Crescent and lightning states exhibit different migration speeds, with crescent being slow and lightning being fast. %The three states differ in their dynamics and display distinct network morphologies that are apparent through visual inspection. 
Regarding network shape, stationary and crescent states form a reticulated network, developing an extended migration front in the crescent state. In the lightning state, however, the network consists of a few thick tubes arranged in a tree-like morphology. To comprehend the benefits of transitioning between network states, we must first answer the question: What is the connection between network morphology and migration dynamics? 

In order to obtain statistics about the morphodynamic behavior, we search for morphometric measures that allow an automatized distinction of the different states directly from the time series of the bright-field images.
We discard measures such as area and perimeter because they are size-dependent. A suitable parameter to describe the mass distribution within a shape is the radius of gyration $R_{\mathrm{gr}}$. To compare networks of different sizes, $R_{\mathrm{gr}}$ is scaled by the square root of the area of the organism, $\sqrt{A}$ (SI Appendix, Fig.~S1). The time series of the $R_{\mathrm{gr}}/\sqrt{A}$ first indicates the variability of the morphological states (Fig.~\ref{fig:phaseplot}E and SI Appendix, Fig.~S2 and~S3) with higher values in the lighting state (blue background). In comparison, smaller values coincide with the stationary state (yellow background).
While the scaled radius of gyration measures the biomass distribution within the plasmodium, the shape of the network is evaluated by the solidity, given as $A/A_c$, where $A_c$ is the area inside the convex hull, the smallest convex perimeter outlining the network (SI Appendix, Fig.~S2 and~S3). 
To assess variations in migration dynamics, we compute the mean migration front velocity $v_{\mathrm{front}}$ for each time point for each plasmodial network. The first obvious choice for velocity, the velocity of the center of mass of the network, fails here because it is susceptible to network reorganization without displacement, such as pruning of individual network tubes (SI Appendix, Fig.~S2 and~S3).\\
\newline
After quantifying the scaled radius of gyration $R_{\mathrm{gr}}/\sqrt{A}$ as a biomass distribution parameter, the solidity $A/A_c$ as a shape measure, and the mean migration front velocity $v_{\mathrm{front}}$ for all plasmodia, we train a neural network by providing a set of  training data that has been visually classified by a human to distinguish between the different states. Note that instead of training the AI to evaluate gray-scale image features directly, we provide the three parameters as training data, resulting in a more accurate phase space that covers the wide range of transitions during network reorganization.  We find that in the stationary state, $v_{\mathrm{front}}$ and $R_{\mathrm{gr}}/\sqrt{A}$ are consistently small, while $A/A_c$ is close to one (Fig.~\ref{fig:phaseplot}F/G). The values of $v_{\mathrm{front}}$ and $R_{\mathrm{gr}}/\sqrt{A}$ increase for both migratory states and are largest for lightning. $A/A_c$ decreases and drops to 0.2 for the lightning state. The standard deviations of the means overlap due to the sheer range of morphological transitions captured in the data points. Note that the organism's size only marginally affects the measures, which are also independent of the environment (SI Appendix, Fig.~S4), making them robust criteria for identifying the three morphological network states.\\
\newline
Individual network states, such as lightning with its tree-like morphology, are reminiscent of theoretically predicted optimal morphologies that minimize transport costs due to viscous energy dissipation of the fluid flow at a fixed building cost~\citep{katifori_Damage_2010,corson_Fluctuations_2010}.
However, in this context, a reticulated network morphology, as observed in the stationary state, would be formed if the building cost is lower than for a time point with the lightning state. To solve the puzzle of how network states that are theoretically optimal under different cost constraints can be interchangeably swapped within a single organism, we next turn to quantify the associated costs of network states.

%%%%%%%%%%%%%%%%%%%%%%%%%%%%%%%%%%%%%%%%%%%%%%%%%%%%%%%%%%%%%%%%%%%%%
%-------------------------------------------------------
% ENERGY BUDGETING
%-------------------------------------------------------
\subsection{\label{sec:part2} Network morphology states show trade-off between transport and building costs}
To quantify the energetic cost associated with the three network states, we follow the large body of theoretical work on optimal flow networks~\citep{murray_Physiological_1926,katifori_Damage_2010,corson_Fluctuations_2010, ronellenfitsch_Phenotypes_2019,hu_Adaptation_2013}, known to also capture out-of-equilibrium states~\citep{akita_Experimental_2017}. The total power of operating a tube segment in flow networks results in a sum of transport costs, $\mathcal{C}_{\mathrm{trans}}$, measured by the viscous energy dissipation of the fluid flow, and the building cost $\mathcal{C}_{\mathrm{build}}$~\citep{murray_Physiological_1926}, the metabolic cost of the tube segment related to the tube volume. 
%%%%%%%%%%%%%%%%%%%%%%%%%%%%%%%%%%%%%%%%%%%%%%%%%%%%%%%%%%%%%%%%%%%%%
%_------------------FIGURE---------------------------
\begin{figure}[ht!]
\includegraphics[width=0.49\textwidth]{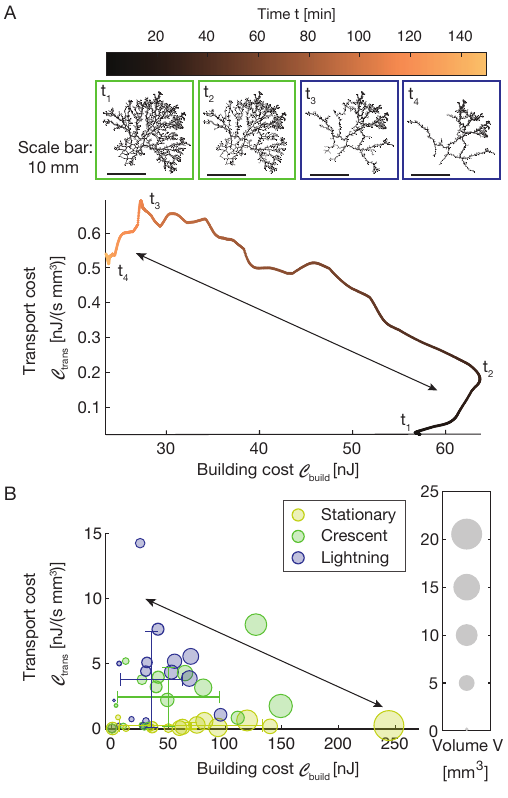}
\caption{\label{fig:tradeoff} Change in network morphology associated with trade-off between transport cost $\mathcal{C}_{\mathrm{trans}} = \partial_t E_{\mathrm{v}}/V$ and building cost $\mathcal{C}_{\mathrm{build}} = E_{\mathrm{el}}$. (A) The high-time resolution data set is a spatially stable individual, allowing for time-resolved extraction of costs. The plasmodium reduces building cost by increasing transport cost when transitioning from crescent ($t_1$, $t_2$) to lightning state ($t_3$, $t_4$). (B) Trade-off between building and transport cost recaptured in large statistics of low-time resolution network morphologies of migrating \textit{P.~polycephalum}. Error bars indicate the standard deviation of the distribution of data sets in each state.}
\end{figure}
To account for flow velocity scaling with overall network size in \textit{P.~polycephalum}~\cite{kuroda_Allometry_2015}, we normalize the overall energy dissipation of an organism by the organism volume, defining transport cost as: 
\begin{equation*}
    \mathcal{C}_{\mathrm{trans}}=\frac{\partial E_{\mathrm{v}}}{\partial t}\frac{1}{V}=\left\langle\sum_{k=1}^N \frac{Q_k(t)^2}{C_k}/\sum_{k=1}^N V_k\right\rangle_t,
\end{equation*} 
where we sum over a network consisting of $k\in~N$ cylindrical tubes with tube volume $V_k~=~\pi a_{0, k}^2 l_k$ for tubes of individual length $l_k$ and base radius $a_{0,k}$. Energy dissipation in tube~$k$ is determined by the volumetric flux $Q_k(t)=\pi a_{0, k}^2 \bar{u}_k(t)$ and hydraulic conductance $C_k = \pi a_{0, k}^4/8\mu l_k$, where $\bar{u}_k(t)$ denotes the cross-sectionally averaged flow velocity. Thus, viscous energy dissipation is directly accessible from measurements of the network architecture, $\{a_{0, k}, l_k\}$, cytoplasm viscosity $\mu=0.275$~Pa~s~\cite{oettmeier_Lumped_2019, sato_Rheological_1983}, and calculations of flows $\bar{u}_k(t)$ driven by tubular contractions from the base radius, $a_k(t)-a_{0,k}$~\citep{marbach_Vascular_2023,marbach_Pruning_2016} which we average over two contraction periods.
Considering the rhythmic tubular contractions as key metabolic costs, we quantify \textit{P.~polycephalum}'s building costs by the network's time-averaged elastic energy~\citep{bauerle_Living_2020},
\begin{equation*}
    \mathcal{C}_{\mathrm{build}}=E_{\mathrm{el}} =\left\langle \sum_{k=1}^N \frac{Y\,h_k\pi a_{0,k}}{(1-\nu^2)} l_k\frac{(a_k(t)-a_{0,k})^2}{a_{0,k}^2}\right\rangle_t
\end{equation*}
with $\nu = 1/2$ Poisson's ratio, $h_k=0.1a_{0,k}$ tube wall thickness, $Y=10$ kPa Young's modulus~\citep{naib-majani_Morphology_1988, fessel_Indentation_2018, norris_Elasticity_1940} also time-averaged over two contraction cycles. Note that, in this case, the elastic energy as the specific choice of building cost allows a direct quantification while maintaining the original geometric scaling of building costs~\citep{murray_Physiological_1926}.
As we focus on network cost associated with the different network states emerging on the scale of hours, we average both transport and building costs over \textit{P.~polycephalum's} intrinsic time scales~\citep{marbach_Vein_2023}.\\

The network architecture, $\{a_{0, k}, l_k\}$, tube dynamics $a_k(t)$, and resulting flow velocities $\Bar{u}_k(t)$ are retrieved from data in two steps. First, we follow contraction dynamics in the network with high time resolution acquired with a 6-second interval, transitioning from crescent state, see time points $t_1$ and $t_2$ in Fig.~\ref{fig:tradeoff}A, to lightning state, see time points $t_3$ and $t_4$ in Fig.~\ref{fig:tradeoff}A (Movie~S3). Due to its immobility, the high-time resolution network allows us to extract tube dynamics $a_k(t)$ from bright-field images. The cross-sectionally averaged flow velocity $\bar{u}_k(t)$ follows from solving for laminar flow and conservation of fluid volume at each node in the network~\citep{marbach_Vein_2023, marbach_Vascular_2023}. 
Migration fronts that develop during the transition only play a passive role in the overall flow and contractility~\citep{oettmeier_Form_2018} and are, therefore, not included in the analysis.   

The quantification of costs reveals that the crescent state comes at a high building cost due to its reticulated network structure; however, the ensuing flows are at low velocities from the diversion among many competing routes, resulting in small transport costs. As the network reorganizes into a lightning state, mass is relocated into fewer, bigger tubes, reducing the necessary building costs. Simultaneously, flow velocities and thus dissipation, aka transport cost, increase. Noteworthy, fluctuations in contractions decrease during the transition from crescent to lightning (SI Appendix, Fig.~S8) in line with theoretical predictions on the impact of fluctuations during cost optimization~\cite{ronellenfitsch_Phenotypes_2019}.

The high-time resolution data set already suggests that building and transport cost variations are traded off as networks reorganize. 
We next turn to quantify both costs in the large statistics of the low-time resolution data sets underlying Fig.~\ref{fig:phaseplot} and employ the high-time resolution network data to confirm that direct flow calculations from measured tube dynamics agree with flow predictions based on modeling contractions as a peristaltic wave of 20\% contraction amplitude along the network's longest axis following Ref.~\cite{alim_Random_2013,marbach_Pruning_2016} (SI Appendix, Fig.~S6), which we use to quantify flow velocities based on extracted network morphologies in the low-time resolution data.
Pooling networks by size and state reveals that building cost is largest for stationary states, about double the cost for crescent states, which is twice the building cost for lightning states, for each respective plasmodia size, see Fig.~\ref{fig:tradeoff}B. Transport costs show the opposite behavior: the lightning state is the most costly, with twice the amount of dissipation than crescent states, itself double the amount of transport cost as the stationary state. Note that while transport costs appear to drop towards zero in the stationary case, they are, in fact, merely an order of magnitude smaller (see log-scale SI Fig.~S7). Building and transport costs appear to be in a trade-off facilitated by a network morphology change. Note that contrary to the theoretical hypothesis, dissipation is not always minimized. Which advantage does \textit{P.~polycephalum} gain from investing in higher flows by reorganizing network morphology?
%%%%%%%%%%%%%%%%%%%%%%%%%%%%%%%%%%%%%%%%%%%%%%%%%%%%%%%%%%%%%%%%%%%%%
% %------------------------------------------------------------------
% % DIFFERENT SEARCH
% %------------------------------------------------------------------
\subsection{\label{sec:part4} Search strategies vary with network state}
%%%%%%%%%%%%%%%%%%%%%%%%%%%%%%%%%%%%%%%%%%%%%%%%%%%%%%%%%%%%%%%%%%%%%
%
%
%_------------------FIGURE---------------------------
\begin{figure}[ht!]
\includegraphics[width=0.49\textwidth]{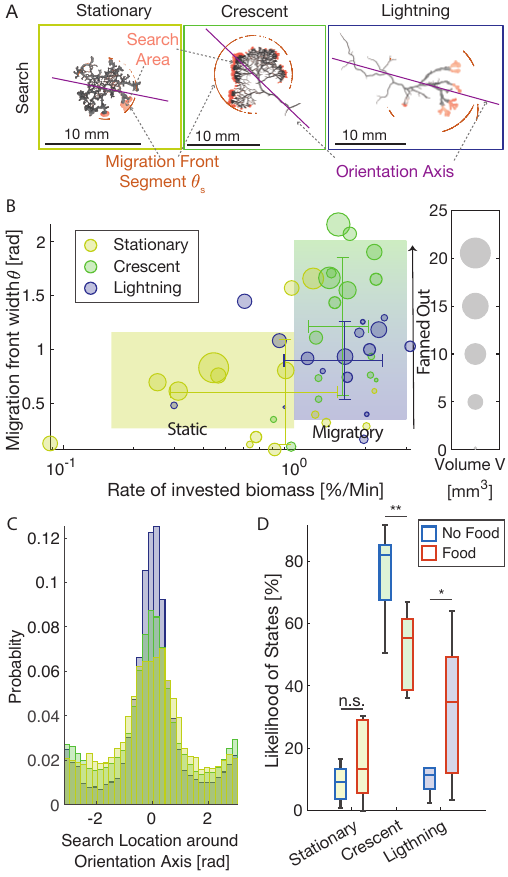}% Here is how to import EPS art
\caption{\label{fig:function} Different network states in \textit{P.~polycephalum} allow for a variety of search strategies. (A) Search in \textit{P.~polycephalum} is arising from migration front dynamics. (B) The rate of invested biomass separates different network states into static and migratory. Migratory states differ in how much their fronts are fanned out, quantified by migration front width. (C) \textit{P.~polycephalum} searches along the network orientation axis for lightning and distributes search broader for crescent and stationary. (D) \textit{P.~polycephalum} adapts the likelihood to spend time in different morphodynamical states according to its environment. p-values are gained by standard null hypothesis significance testing: * $p\leq0.05$ and ** $p\leq0.01$.}
\end{figure}
We quantify each state's search dynamics with three parameters to investigate the potential diversity in each network state's functionality. The migration front behavior governs search dynamics (Fig.~\ref{fig:function}A). First, we infer the overall biomass allocated into migration fronts per minute by quantifying the intensity of the transmitted light in the migration front area compared to the total area covered by the organism, invoking Beer-Lambert law. Second, we investigate \textit{P.~polycephalum}'s strategy in scanning its local environment by calculating the width of the migration front $\theta$. Here, we map the boundary pixels of the migration front onto a unit circle and sum over the angular width of each individual migration front segment $\theta_s$. Third, we introduce the search location relative to the orientation of the plasmodium, given by the main axis of the network's radius of gyration (Fig.~\ref{fig:function}A). Aligning each network's intrinsic orientation axis to~0$^\circ$, we determine the relative angular position of the migration front boundary pixels and combine all data sets state-wise into the separate histograms in Fig.~\ref{fig:function}C.

The rate of invested biomass separates network states into migratory (crescent and lightning state) versus static (stationary state) (Fig.~\ref{fig:function}B). Invested biomass is one magnitude smaller in the stationary state than in both migratory states. Although plasmodia in the lightning state are twice as fast as plasmodia in the crescent state, the rate of invested biomass is the same. Differences among both migratory states only appear when quantifying the environmental scanning strategy. The migration front width is only half as wide in the lightning state as in the crescent state. The invested biomass is, therefore, used to achieve high velocities in the lightning state and to scan a broad environment in the crescent state. As the rate of invested biomass is low for stationary networks, the migration front width is also comparably small. With a low velocity in the stationary state, \textit{P.~polycephalum} invests all of its biomass in local environment scanning. 
Mapping out the search location around the orientation axis of the plasmodium further reveals that both stationary and crescent plasmodia tend to search everywhere around the organism, which is apparent through the wide distribution of front locations (Fig.~\ref{fig:function}C). In the lightning state, however, the search is very focused in the immediate direction of the plasmodium, visible in the very narrow front location distribution.
From lightning via crescent to stationary state, the search location distribution widens. Search dynamics in the different network states vary from unidirected search in the stationary and crescent states to directed search in the lightning state. Combining all three parameters characterizing search dynamics thus reveals that each network state represents a unique search strategy.

To probe if different search strategies are an inherent function of the network-forming forager's environment, we quantify the likelihood of different states in environments that are either plain or scattered with patchy food sources. 
We find that food-encountering \textit{P.~polycephalum} is twice as likely to be in the lightning or stationary state (stationary:crescent:lightning:20\%:50\%:30\%) compared to non-successful foragers (stationary:crescent:lightning:10\%:70\%:15\%), see Fig.~\ref{fig:function}D. We deduce that continuous network reorganization enables richer search strategies that are impacted by the forager's environment and, thus, environmental changes.

\section{Discussion}\label{sec:discussion}
Following network reorganization in \textit{P.~polycephalum} for days, we identify a transition between three different network states unique in morphology and migration dynamics. We estimate the energetic cost associated with each state and reveal that morphological variability trades off transport and building costs. Quantifying each state's search strategy reveals that dynamic network reorganization within the trade-off of costs allows the network-forming forager to continuously switch between strategies. While different environments do not affect the morphological variability, they drive a network-forming forager to change the search dynamics. 

Network reorganization is a general optimization process not solely found in network-forming foragers but also developing blood vasculature~\citep{papadopoulos_Comparing_2018, chen_HaemodynamicsDriven_2012}, bile canalicular networks~\citep{dahl-jensen_Deconstructing_2018}, and leaf venation~\citep{katifori_Damage_2010}. However, those networks evolve into a single network morphology in line with the minimization of energetic costs predicted by theoretical models at a fixed ratio of transport versus building cost~\citep{murray_Physiological_1926, kirkegaard_Optimal_2020,katifori_Damage_2010,corson_Fluctuations_2010, ronellenfitsch_Phenotypes_2019,hu_Adaptation_2013}. Our estimates of the energetic costs in \textit{P.~polycephalum}'s network states suggest that the allocation of transport versus building also constrains network morphology. Nevertheless, instead of optimizing for a single network morphology, varying energy allocation over time enables \textit{P.~polycephalum} a dynamic range of network states and search strategies.

Different network states in \textit{P.~polycephalum}, termed stationary, crescent, and lightning by us, align with prior observations of the crescent state in chemotaxis~\citep{westendorf_Quantitative_2018} or network coarsening~\citep{baumgarten_Dynamics_2014}. The transition from stationary, isotropic networks to migratory, crescent networks has been described as a change from exploiting a food source (stationary) to exploring new sources (crescent)~\citep{halvorsrud_Growth_1998, vogel_Transition_2017, latty_Food_2009}. Our findings here broaden the diversity of migratory states, now accounting for lightning in addition to crescent, showing that network reorganization allows for rich search dynamics in network-shaped foragers compared to other species~\citep{lee_Novel_2018,bartumeus_Optimal_2009,viswanathan_Physics_2011}. 
Network adaptation from sheet-like (stationary or crescent) to more tree-like (lightning) structures also results from environmental variations~\citep{takamatsu_Environmentdependent_2009, nakagaki_Interaction_2000}. However, while we observe that food encounters change the likelihood of state transitions, they are not the sole trigger; transitions occur even without visible environmental stimulation and happen frequently. This internal trigger might enhance foraging success, which warrants further investigation. 

Previous studies on the search strategies of \textit{P.~polycephalum} involved inoculating plasmodia directly onto food sources, demonstrating that plasmodial networks fully explore their environments before optimizing transport between large food sources~\citep{nakagaki_Mazesolving_2000, tero_Rules_2010, tero_Mathematical_2007}. In contrast, our experiments use plasmodia with limited biomass and small food sources, necessitating a trade-off between optimizing transport and minimizing network-building costs. Note, that \textit{P.~polycephalum} additionally optimizes its foraging strategy by leaving behind a slime trail as an external memory~\citep{reid_Slime_2012}, adjusting its diet based on food patch composition~\citep{dussutour_Amoeboid_2010}, and modulating its oscillation patterns to store environmental information~\citep{boussard_Adaptive_2021, fleig_Emergence_2022}. In this context it is noteworthy that we also find transitions between different morphological states to be linked to changes in the contraction patterns, suggesting oscillation dynamics to be linked to the optimization process in \textit{P.~polycephalum}.

Not only \textit{P.~polycephalum}'s network reorganization dynamics but also the observed network states resemble foraging fungi~\citep{bebber_Biological_2007, glass_Hyphal_2004, boddy_Saprotrophic_1999}, suggesting that the cost constraints and search strategy dynamics found here might well be independent of the biological make-up. As we find network states to be tied to biomass reallocation, adapting the network state population to environmental conditions gives a first insight into how network-forming foragers might re-route resource flows in changing environments.

%-----------------------------------------------------------------
% METHODS
%-----------------------------------------------------------------
\section{Methods}\label{sec:methods}

\subsection{Culturing and imaging of \textit{P.~polycephalum}}
\noindent\textit{P.~polycephalum} networks (Carolina Biological Supplies) were inoculated from microplasmodia grown from liquid culture~\citep{daniel_Pure_1961, daniel_Hematinrequiring_1962} onto 1.5\% (w/v) nutrient-free agar. Half of the plasmodia are presented with localized patches of heat-killed E.~coli. The food sources are distributed in patterns varying from randomly distributed to circle- and spiral-like without guaranteeing a successful foraging of the plasmodia.
For the low-time resolution data sets, experiments were started 1-5~h after inoculation and carried out for 24-60~h. Images of the networks were acquired every 5 or 10 Min with a Zeiss Axio Zoom V.16 microscope equipped with a Hamamatsu ORCA-Flash 4.0 digital camera and a Zeiss PlanNeoFluar 1×/0.25 objective with a spatial resolution of 80-120~px/mm. A green filter (550/50 nm) was placed over the transmission light source of the microscope to diminish \textit{P.~polycephalum's} response to the light. A custom-designed top-stage incubator from Okolabs or Pecon controlled the temperature and humidity of the experimental environment.
For the high-time resolution data, spatially stable \textit{P.~polycephalum} networks were trimmed from a well-established network and left to relax for 30 min before the onset of imaging to recover from trimming. Using the Zeiss Axio Zoom V.16 setup again here, an image was taken every 6~s for 3~h.% as the plasmodium started migrating afterward.
\subsection{Morphodynamic analysis}
One bright-field image frame contains 25 individual image tiles each. All tiles were converted into 8-bit tiff-files using Zeiss Zen 2. A rolling-ball algorithm removed the background on each tile. The Microscopy Image Stitching Tool~\citep{blattner_Hybrid_2014, chalfoun_MIST_2017} stitched the tiles into one frame. 
\textit{P.~polycephalum} networks were extracted with a custom-written MATLAB (The MathWorks) code, creating a binary image by intensity thresholding and closing single pixels. Using the MATLAB built-in functions \textit{bwconncomp} and \textit{regionprops}, morphological features like network area  $A$, convex area size $A_c$, bounding box, centroid, and perimeter, among others were extracted. The location and intensity of each pixel in the binary image were stored.   
Beer-Lambert's law allowed a linear relation between the gray-value intensity and the biomass of the translucent slime mold. 
For the radius of gyration, hence, all pixels $n$ in the binary image were weighted by their gray-scale value $f_{n}$. The radius of gyration is given by a summation over all pixels and their squared distances $dx_n$ and $dy_n$ with respect to the network's center of mass, respectively,
\begin{equation*}
    R_{\mathrm{gr}} = \sqrt{\frac{\sum_{\mathrm{pixel n}} (dx_n^2 + dy_n^2)\cdot f_n}{f_{\mathrm{tot}}}},
\end{equation*}
divided by the total plasmodium's gray-scale intensity $f_{tot}$ as a measure of relative biomass. The migration velocity of a growth front was extracted by mapping out the distance between growth front lines of consecutive frames~\citep{westendorf_Quantitative_2018}. 
Morphology and velocity data enabled a visual distinction of three different morphological states. 
A neural network with MATLAB \textit{Deep Learning Toolbox} was constructed to classify the morphodynamic state of the network frame by frame. 
We used each frame's scaled radius of gyration $R_{\mathrm{gr}}/\sqrt{A}$, solidity $A/A_c$, and migration front velocity $v_{\mathrm{front}}$ as input data. A single observer visually inspected and categorized ~25\% of the frames of several data sets, resulting in about 1050 human-labeled frames, to generate the training data. As the available data was limited, training and validation data were constructed by splitting the input data into several batches of 32 frames each, first used to validate previously trained batches and then included in the training data. 
The training data passed through a sequence input layer with 3 nodes into a series of fully connected and leaky ReLU (Rectified Linear Unit) layers with 100 nodes each. ReLU layers perform threshold operations on the data~\citep{nielsen_Neural_2015}. In the fully connected layer, each neuron connects to each neuron from the previous layer. The input data is weighted and gets an added bias, preventing the neural network from down-sampling the data. The last layer is a softmax layer with 3 nodes with a normalized exponential applied to the input. It serves as a variation of the logistic sigmoid function, enabling the reduction of weights to a fixed classification~\citep{bishop_Pattern_2006}. As an output, each frame was categorized into one of the three visually observed states, enabling further analysis of the functionality of these states. 
A training cycle contains of 300 epochs with 9900 iterations in total, leading to an accuracy of 95~\% in the state detection.
\subsection{Flow extraction from spatially stable network data}
Flows were extracted from recorded contractions of a high-time resolved network to quantify the dissipation cost arising from the cytoplasmic flows. A custom-developed MATLAB code quantified network morphology and dynamics following Ref.~\citep{bauerle_Spatial_2017}: single frames were binarized by thresholding to determine the network’s structure, using pixel intensity as well as pixel variance information extracted from an interval of images around the processed image. Binarized images were skeletonized with the smallest resolved structures of 1.5-3~px (see SI Appendix, Fig.~S5). Tube radius and the corresponding intensity of transmitted light were measured along the skeleton. The two quantities are correlated according to Beer-Lambert’s law. Radius variations within tube segments were smoothed by surface fitting with the MATLAB function \textit{gridfit}~\citep{derrico_Surface_2023}. A custom-developed MATLAB code calculated flows within tubes from the extracted network structures based on the conservation of mass following Ref.~\citep{marbach_Vein_2023}: A discretized dynamic network structure was generated and mapped onto the first skeleton.
The network was over-discretized in time by adding two linearly interpolated values between each frame to guarantee a more accurate flow calculation. Based on Ref.~\citep{alim_Random_2013}, considering Kirchhoff laws and Poiseuille flow for each node in the network, flow, and pressure in each segment were calculated. As detected from sequential images, the actual live contractions $a_k(t)$ were used as input here. Compared to Ref.~\citep{marbach_Vein_2023}, the data set was split into 100-time step segments each to account for biomass flowing into migration fronts. For 100 time steps, the cytoplasmic mass is conserved, adjusting the radius $a_k(t)$ to ensure that Kirchhoff's laws were solved with good numerical accuracy. Vanishing tube segments add an additional inflow considered in the calculation. Extracted tube diameters and flows can thus be used to calculate transport and building costs during network reorganization. 

\subsection{Numerical flow simulations in large statistics data}
The low-time resolution data sets presented in Fig.~\ref{fig:phaseplot} taken with a 5 or 10 Min per frame interval do not allow for direct extraction of the contractions of the tubes. Thus, radius dynamics were numerically modeled onto each frame to predict flow and tube dynamics. To incorporate migration fronts, we expanded on the custom-developed MATLAB code of Ref.~\citep{bauerle_Spatial_2017} by now reducing the gray-scale image frames to a skeleton of the network: On the gray-scale image of the region of interest defined by the binarized image a gray weighted distance transform~\citep{qian_Gray_1999} is produced using the built-in MATLAB function \textit{graydist}. The gray-distance map was used to enhance features such as growing tubes in the fronts, which, combined with a thinning process, led to the inclusion of these features in the skeleton. From the skeleton, intensities and diameters were extracted as described previously. 
Each image frame was then used as an individual base structure with its base radius $a_{0,k}$ to map a peristaltic wave onto the extracted network, predicting contractions and flow patterns for a selected time over 2000 iterations~\citep{marbach_Pruning_2016}.\\

%\subsection{Energetic cost estimation}
%From the retrieved radial contractions and resulting flows, we can assess the energetic costs for building the network and transport within the network. Transport costs were measured by the viscous energy dissipation $\partial_t E_{v}$ in a network consisting of $k\in N$ cylindrical tubes of individual length $l_k$, pervaded by a flow of cross-sectionally averaged velocity $\Bar{u}_k(t)$. Transport costs hence resulted in
%\begin{align}
%    \partial_t E_{\mathrm{v}} =\biggl\langle\sum_k^N 8\pi \mu l_k  \Bar{u}^2_k(t)\biggr\rangle_t,
%\end{align}
%where $\mu$ denotes the fluid's viscosity. 
%We averaged the energetic costs over 120~s, to remove fluctuations driven by the contractions. Furthermore, \textit{P.~polycephalum} shows an intrinsic period of roughly 20~Min. Hence, we detrended the data over 20~Min according to Ref.~\citep{marbach_Vein_2023}.
%% 
%
\subsection{Search dynamics analysis}
The search behavior of \textit{P.~polycephalum} is strongly coupled to the dynamics of the migration front, which we extracted when calculating the migration front velocity. A difference between two different masks gave the searched area.
The different time resolution of the data sets require an adaptation of the time interval studied for the search, which was set to 10~Min intervals (the maximum difference).
$(x,y)_m(t)$-coordinates of the pixels of the search area for each time frame $t$ are centered around the center of mass and then transformed into polar coordinates $(r,\phi)_m(t)$. To remove falsely detected pixels residing from intensity fluctuations and contractions of the tubes in the center of the slime mold network, only pixels with a distance greater than $r_m(t)>0.7 \langle r_s \rangle$ to the center of mass are considered for the analysis. All remaining pixels are sorted according to their angular component. 

Connected migration front segments $s$ are defined as all pixels with less than $0.5\deg$ difference angle between consecutive pixels. 
The migration front width $\theta= \sum_s \theta_s$ is then calculated as the total angular width of each segment $\theta_s \sum_{i=1}^{n-1} (\phi_{i+1}-\phi_i) \mathrm{ with } \{  n\ \epsilon \ \mathrm{s} \}$.
%\begin{align}
%    \theta &= \sum_s \theta_s \\
%    \theta_s &= \sum_{i=1}^{n-1} (\phi_{i+1}-\phi_i) \text{ with } \{  n\ %\epsilon \ \text{s} \}.
%\end{align}
Additionally, we wanted to calculate the location \textit{P.~polycephalum} searched with regard to its orientation axis. All search area pixels $(r,\phi)_g(t)$ are rotated around the orientation of the network $\phi_o$, centering the orientation  at 0~rad. The network's orientation is defined via the main axis of the radius of gyration which can be obtained from the orientation property in the MATLAB built-in function \textit{regionprops}. It is defined as the angle between the horizontal axis and the major axis of an ellipse with the same second moment properties as the organisms shape. The reoriented angular component of the search area pixels then gives the search location.
% \begin{video}
% \href{http://prst-per.aps.org/multimedia/PRSTPER/v4/i1/e010101/e010101_vid1a.mpg}{\includegraphics{vid_1a}}%
%  \quad
% \href{http://prst-per.aps.org/multimedia/PRSTPER/v4/i1/e010101/e010101_vid1b.mpg}{\includegraphics{vid_1b}}
%  \setfloatlink{http://link.aps.org/multimedia/PRSTPER/v4/i1/e010101}%
%  \caption{\label{vid:PRSTPER.4.010101}%
%   Students explain their initial idea about Newton's third law to a teaching assistant. 
%   Clip (a): same force.
%   Clip (b): move backwards.
%  }%
% \end{video}
\section*{Data Availability}
The collected experimental data and code for simulation is available in the mediaTUM repository under https://doi.org/10.14459/2024mp1733537 after publication. 

\begin{acknowledgments}
This work was supported by the Max Planck Society and has received funding from the European Research Council (ERC) under the European Union’s Horizon 2020 research and innovation program (grant agreement No. 947630, FlowMem) to K.A..
We further thank Ron Keuth for the setup of the neural network in Matlab and Leonie Bastin for Movie S3. 
\end{acknowledgments}

% \appendix
% \section{Appendixes}

% \bibliographystyle{abbrv}
%\bibliography{Foraging_Paper_all} % Produces the bibliography via BibTeX.

\end{document}

\title{\large Supplemental Materials: Dynamic cost allocation allows network-forming forager to switch between search strategies
}

\author{Lisa Schick}
\affiliation{%
 Technical University of Munich, Germany: TUM School of Natural Sciences, Department of Bioscience, Center for Protein Assemblies (CPA)
}%
\author{Mirna Kramar}
\affiliation{Institute Curie, UMR168, Paris, France}%Lines break automatically or can be forced with \\
\author{Karen Alim}%
 \email{k.alim@tum.de}
\affiliation{%
 Technical University of Munich, Germany: TUM School of Natural Sciences, Department of Bioscience, Center for Protein Assemblies (CPA)
}%

\date{\today}% It is always \today, today,

\maketitle

\widetext
\section{Calculation of characteristic quantities}
\subsection{\label{secsupp:morphology}Morphological parameters} 
\begin{figure*}[h!]
    \centering
    \includegraphics[width =\textwidth]{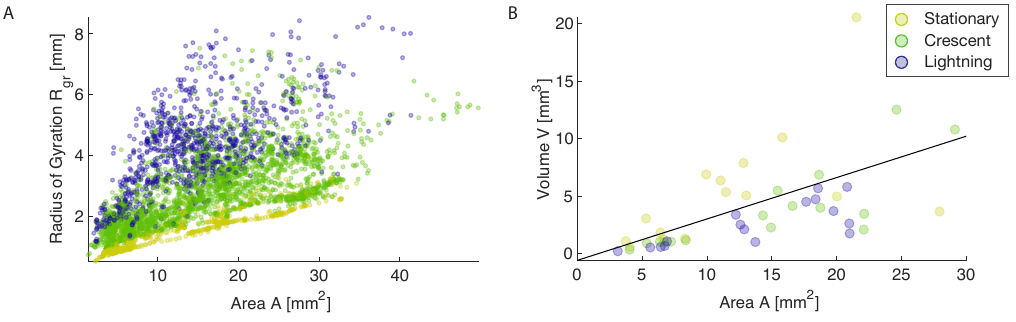}
    \caption{(A) The radius of gyration is size dependent for all frames of the 16 individual \textit{P.~polycephalum} networks. (B) The size parameters volume $V$ and area $A$ have a strong positive correlation (Pearson correlation coefficient $\rho = 0.5196$ and corresponding p-value $p=0$). Here, displayed as an average within the different state for all data set.}
    \label{supp_fig:SizeDep}
\end{figure*}

\begin{figure*}[h!]
   \centering
    \includegraphics[width =0.9\textwidth]{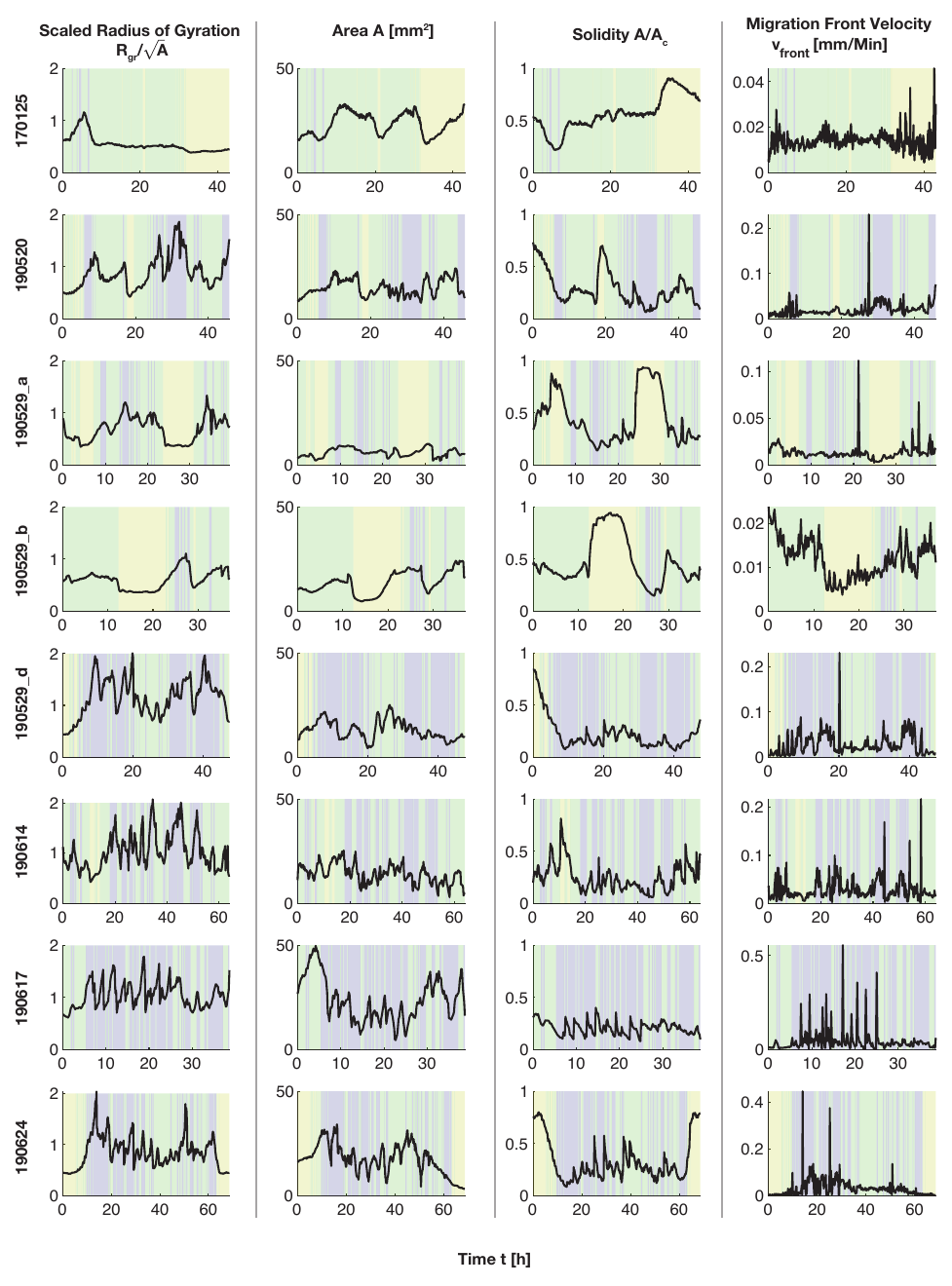} 
    \caption{Time series of the morphology parameters scaled radius of gyration $R_{gr}/\sqrt{A}$, area $A$, solidity $A/A_c$ and migration front velocity $v_{\text{front}}$ for the eight plasmodia with food encounters. The background is highlighted according to the detected morphodynamical state: stationary – yellow; crescent – green; lightning - blue.}
    \label{supp_fig:TimeFood}
\end{figure*}

\begin{figure*}[h!]
   \centering
    \includegraphics[width =0.9\textwidth]{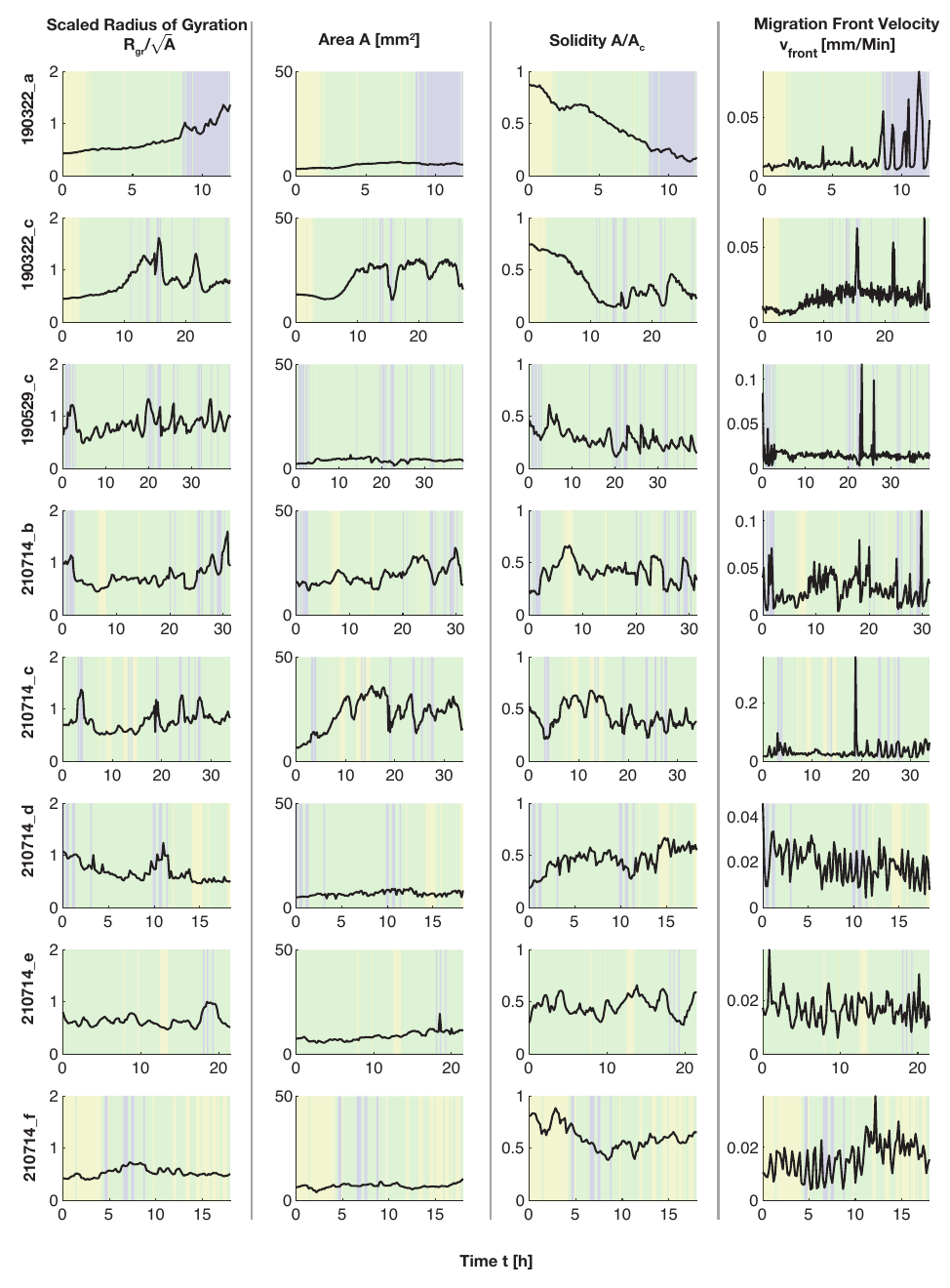} 
    \caption{Time series of the morphology parameters scaled radius of gyration $R_{gr}/\sqrt{A}$, area $A$, solidity $A/A_c$ and migration front velocity $v_{\text{front}}$ for the eight plasmodia without food. The background is highlighted according to the detected morphodynamical state: stationary – yellow; crescent – green; lightning - blue.}
    \label{supp_fig:TimeNoFood}
\end{figure*}

\begin{figure*}[h!]
    \centering
    \includegraphics[width =\textwidth]{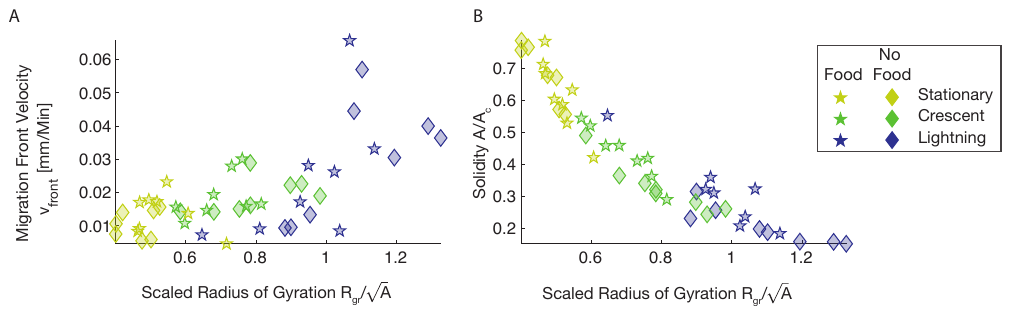}
    \caption{Morphodynamic parameters are independent of the environment displayed similarly to Fig.~1 (A) Migration Front Velocity against Scaled Radius of Gyration averaged over time-steps in the different states for the 16 individual plasmodia. (B) Solidity versus Radius of Gyration also averaged over the individual plasmodia.}
    \label{supp_fig:PhaseFood}
\end{figure*}

Morphological parameters displayed in the main Fig.~1 are calculated for each frame in the single low-time resolution data sets. For a better visualization within the plots, parameters are averaged over all frames in one state within one data set and plotted by the average size of the plasmodium. Size can be described by both, the area $A$ covered by the plasmodium and the volume $V=\sum_k V_k$ given as a sum over the extracted tube volumes $V_k$ (see Fig.~\ref{supp_fig:SizeDep}). The individual time series of the different parameters for all data sets is presented in Fig.~\ref{supp_fig:TimeFood} and~\ref{supp_fig:TimeNoFood}. Categorizing the main Fig.~1 into the plasmodia with and without food encounters, it becomes apparent that the environment does not have an influence on the morphodynamic parameter set (Fig.~\ref{supp_fig:PhaseFood}).
\\
\\
% \subsection*{\label{secsupp:MigrationFront} Migration front dynamics}
% \begin{figure}
%     \centering
%     \includegraphics{Suppl_Figures/SuppFig3.pdf}
%     \caption{Migration front dynamics}
%     \label{fig:migfront}
% \end{figure}
\newpage
\subsection{Spatial resolution of skeletonization}
\begin{figure*}[h!]
    \centering
    \includegraphics[height =5.3cm]{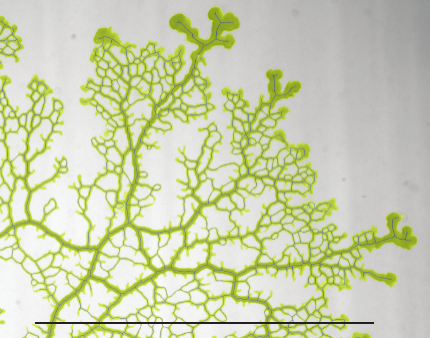}
    \caption{Zoom-in of overlay of binarized image (yellow) and skeleton (blue) on a bright-field image of \textit{P. polycephalum}. Scale~bar:~10~mm}
    \label{supp_fig:SkelRes}
\end{figure*}

The experimentally observed bright-field images are acquired with a spatial resolution of 80-120~px/mm. To skeletonize the bright-field images, at first a binarized image is generated (Fig.~\ref{supp_fig:SkelRes} yellow overlay). The smallest resolved tube diameters of the skeleton (Fig.~\ref{supp_fig:SkelRes} blue overlay) are around 1.5-3~px, i.e. 0.02~mm.
\\
\subsection{\label{secsupp:Flowcomparison}Direct flow calculations vs flow predictions}
\begin{figure*}[h!]
    \centering
    \includegraphics[width =\textwidth]{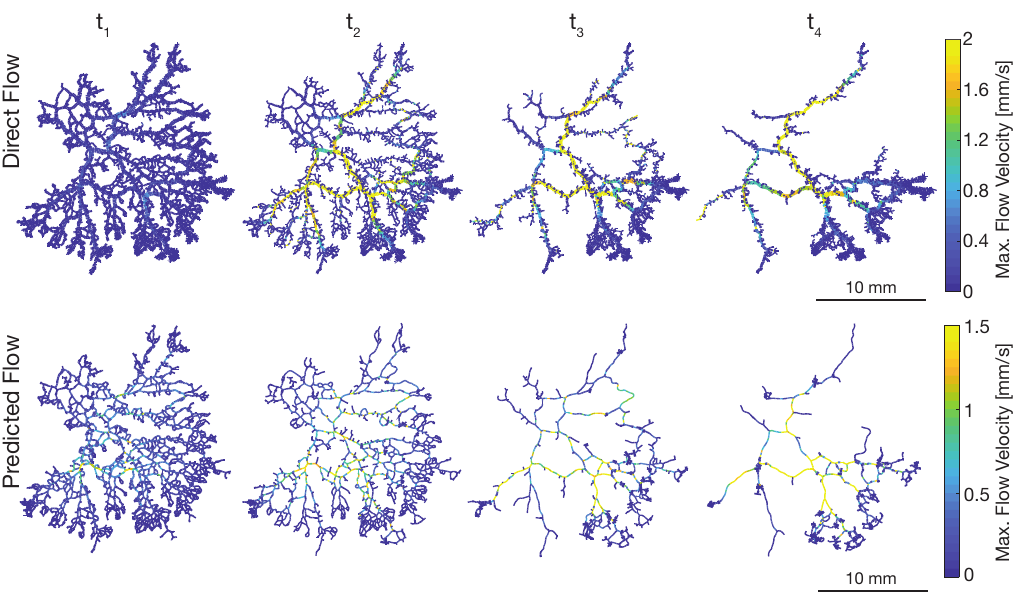}
    \caption{Direct flow calculations (top) from measured tube dynamics agree mostly with flow predictions (bottom) based on modeling contractions as a peristaltic wave of 20\% contraction amplitude along the network's longest axis following Ref.~\citep{alim_Random_2013,marbach_Pruning_2016}.}
    \label{supp_fig:flowcomparison}
\end{figure*}

The high time resolution data set presented in the main Fig.~2A is immobile and allows for a direct flow calculation from the contraction pattern extracted from bright-field images. Applying the contraction modeling as a peristaltic wave of 20\% contraction amplitude along the network's longest axis, which is used for the low-time resolution large statistics data from main Fig.~1 onto the high-time resolution data set, shows that the tube dynamics agree in both cases (Fig.~\ref{supp_fig:flowcomparison}).
\\

\subsection{Magnitude of costs}
\begin{figure*}[h!]
    \centering
    \includegraphics[height = 5.3 cm]{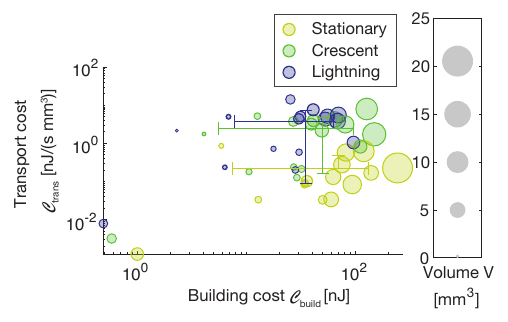}
    \caption{Large magnitude variations highlighted in trade-off between building and transport cost of low-time resolution data sets of migrating \textit{P.~polycephalum}. Error bars indicate the standard deviation of the distribution of data sets in each state.}
    \label{supp_fig:logEnergy}
\end{figure*}

The linear representation of main Fig. 2B shows the trade-off in transport versus building costs in the low-time resolution data of \textit{P.~polycephalum} networks. Fig.~\ref{supp_fig:logEnergy} highlights the different orders of magnitude reached for the transport costs in the different morphodynamical states. 

\section{Fluctuations in contractions}

To quantify the fluctuations in contractions for the crescent and lightning state, we performed Principal Component Analysis (PCA) on the output of the network morphology and dynamics data analysis pipeline of the high-time resolution data. We split the time series into two equal parts, corresponding to the crescent and lightning state, respectively. Using the PCA routine custom-made for~\textit{P. polycephalum} dynamics~\citep{fleig_Emergence_2022}, we obtained the contraction modes and quantified their dynamics for the two foraging states.

\begin{figure*}[h!]
    \centering    
    \includegraphics[width = 0.5\textwidth]{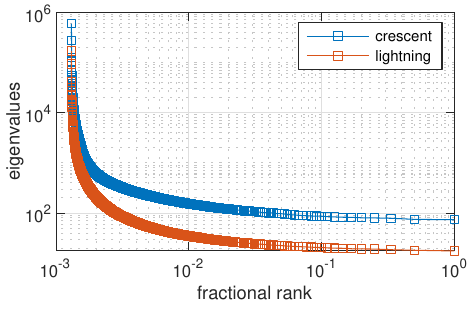}
    \caption{Spectrum of eigenvalues plotted against the fractional rank for the crescent (blue) and lightning state (orange), as output of PCA analysis on the network contractions. }
    \label{supp_fig:PCA_loglog}
\end{figure*}

Figure~\ref{supp_fig:PCA_loglog} shows the mode eigenvalues versus their fractional rank for each of the two states. For the lightning state, the number of eigenvalues (modes) falls off more rapidly than in the crescent state for the same fractional rank. This indicates that a higher number of contraction modes is present in the crescent state, therefore fluctuations in contractions decrease when the network transitions from the crescent to the lightning state.

\bibliography{supplement}